 \documentclass[showpacs,aip,pop,superscriptaddress,groupedaddress]{revtex4}

 \usepackage{graphicx}
 \usepackage{dcolumn}
 \usepackage{bm}
 \usepackage{amssymb}
 \usepackage{amsmath}
 \usepackage{amsfonts}
 \usepackage{color}
 \usepackage{natbib}
 \usepackage{url}
 \usepackage{hyperref}
 \hypersetup{
   colorlinks,
   linkcolor=blue,
   urlcolor=blue,
 }
\graphicspath{{Figures_PoP19/}}

\newcommand{\refsec}[1]{\ref{sec:#1}}

\begin{document}

 \title{Non-linear modeling of the threshold between ELM mitigation and ELM suppression by Resonant Magnetic Perturbations in ASDEX Upgrade}
 \author{F.Orain} \affiliation{\emph{CPHT, Ecole Polytechnique, CNRS, 91128 Palaiseau, France}} \affiliation{\emph{Max-Planck-Institute for Plasma Physics, 85748 Garching, Germany}}
 
 \author{M.Hoelzl} \affiliation{\emph{Max-Planck-Institute for Plasma Physics, 85748 Garching, Germany}}
 \author{F.Mink} \affiliation{\emph{Max-Planck-Institute for Plasma Physics, 85748 Garching, Germany}}
 \author{M.Willensdorfer} \affiliation{\emph{Max-Planck-Institute for Plasma Physics, 85748 Garching, Germany}}
 \author{M.B\'ecoulet} \affiliation{\emph{CEA, IRFM, 13108 Saint-Paul-Lez-Durance, France}}
  \author{M.Dunne} \affiliation{\emph{Max-Planck-Institute for Plasma Physics, 85748 Garching, Germany}}
 \author{S.G\"unter} \affiliation{\emph{Max-Planck-Institute for Plasma Physics, 85748 Garching, Germany}}
 \author{G.Huijsmans} \affiliation{\emph{CEA, IRFM, 13108 Saint-Paul-Lez-Durance, France}} \affiliation{\emph{Eindhoven University of Technology, 5600 MB Eindhoven, The Netherlands}}
  \author{K.Lackner} \affiliation{\emph{Max-Planck-Institute for Plasma Physics, 85748 Garching, Germany}}
 \author{S.Pamela} \affiliation{\emph{CCFE, Culham Science Centre, Abingdon, Oxon, OX14 3DB, UK}}
  \author{W.Suttrop} \affiliation{\emph{Max-Planck-Institute for Plasma Physics, 85748 Garching, Germany}}
  \author{E.Viezzer} \affiliation{\emph{Dept. of Atomic, Molecular and Nuclear Physics, University of Seville, 41012 Seville, Spain}}
 \author{the ASDEX Upgrade team} \affiliation{\emph{Max-Planck-Institute for Plasma Physics, 85748 Garching, Germany}}
 \author{the EUROfusion MST1 team} \affiliation{\emph{See author list in H. Meyer \textit{et al.}, \textit{Nucl. Fusion} 57, 102014 (2017)}}

% \date{\today}

 \begin{abstract}
The interaction between Edge Localized Modes (ELMs) and Resonant Magnetic Perturbations (RMPs) is modeled with the magnetohydrodynamic code JOREK using experimental parameters from ASDEX Upgrade discharges. According to the modeling, the ELM mitigation or suppression is optimal when the amplification of both tearing and peeling-kink responses result in a better RMP penetration. The ELM mitigation or suppression is not only due to the reduction of the pressure gradient, but predominantly arises from the toroidal coupling between the ELMs and the RMP-induced mode at the plasma edge, forcing the edge modes to saturate at a low level. The bifurcation from ELM mitigation to ELM suppression is observed when the RMP amplitude is increased. ELM mitigation is characterized by rotating modes at the edge, while the mode locking to RMPs is induced by the resonant braking of the electron perpendicular flow in the ELM suppression regime.
 \end{abstract}


 \maketitle

\section{Introduction}

Research in magnetically-confined fusion plasmas mainly focuses on tokamak and stellarator devices. In the ITER tokamak under construction, the high-confinement regime (or ``H-mode") has been chosen as operating regime, since it is characterized by an improved confinement induced by a transport barrier at the edge \cite{ITER_review}. This transport barrier is associated with a large edge pressure gradient raising the whole pressure profile on the so-called ``pedestal". However, instabilities named Edge Localized Modes (ELMs), observed in tokamaks and sometimes in stellarators \cite{Zohm_ELM_review}, are triggered beyond a certain threshold in pressure gradient (ballooning modes) and/or plasma current density (peeling modes). ELMs induce the quasiperiodic relaxation of the plasma edge which leads to the expulsion of a large particle flux burst and the deposition of a large transient heat flux on plasma-facing components. The ELM-induced heat flux might be intolerably large for the ITER divertor targets \cite{ITER_review}, thus reliable methods to control or suppress ELMs must be mastered. 

A promising control method consists in applying non-axisymmetric magnetic perturbations, smaller than the toroidal magnetic field by a factor of around $10^4$. These perturbations are aimed to induce magnetic reconnection on the resonant surfaces, characterized by a safety factor $q=m/n$, where $m$ and $n$ are respectively the poloidal and toroidal mode numbers. They are therefore called Resonant Magnetic Perturbations or RMPs. The original goal of RMPs, coming from Tore Supra's ergodic divertor \cite{Ghendrih_PPCF96}, is to create magnetic island chains on the resonant surfaces located at the plasma edge. If the islands are sufficiently large, they overlap and produce a chaotic (or stochastic) layer: in this case, the perpendicular transport is increased and the pressure gradient is therefore reduced. The initial aim of the RMPs is thus to slightly reduce the pressure gradient just below the ELM-triggering threshold, while keeping the pressure large enough to maintain a good confinement. RMPs successfully managed to fully suppress ELMs in DIII-D, ASDEX Upgrade, KSTAR and EAST \citep{Evans_PRL, Suttrop_PPCF17, Jeon_PRL12, Sun_PRL} and to mitigate them in JET and MAST  \citep{Liang_PRL07, Kirk_PRL12}. However the physics of the ELM mitigation and suppression is actually more complicated than this simple picture, in particular because the plasma response to the RMP application is capable to make them ineffective. 

The plasma response, intensively studied theoretically \citep{Fitzpatrick, Heyn08, Nardon_NF10, Liu_PoP10, Yu_NF11, Becoulet_NF12, Ferraro_PoP12, Kaveeva_NF12, Orain_PoP13}, is understood as follows. The electron perpendicular flow generate currents on resonant surfaces, which can induce a magnetic field opposite to the applied perturbation, resulting in a zero net perturbation penetrating in the plasma: RMPs are then screened. More recently, it was found that RMPs can also excite marginally stable peeling-kink modes at the plasma edge, capable to improve the RMP penetration \citep{Paz-Soldan_PRL15, Ryan_PPCF15, Orain_NF17}. 

This paper describes the impact that both ``resonant" and ``peeling-kink" responses have on the ELM mitigation and suppression. The interaction between RMPs and ELMs was modeled with the magnetohydrodynamic (MHD) code JOREK \cite{Huysmans_NF07} including two-fluid diamagnetic and neoclassical effects \cite{Orain_PoP13}. In section \ref{sec:sec2}, the generic features of the ELM control by RMPs, obtained in modeling in close comparison with the experimental observations of ASDEX Upgrade discharges, are presented. In section \ref{sec:sec3}, the threshold between ELM mitigation and ELM suppression is modeled above a given RMP amplitude: the mechanism inducing the ELM suppression is described. Last, the conclusion and a discussion are provided in section \refsec{conclu}.

\section{ELM control by Resonant Magnetic Perturbations in ASDEX Upgrade: generic features} \label{sec:sec2}

This section describes the main mechanisms determining how the plasma response affects the coupling between ELMs and RMPs in the non-linear simulations.

\subsection{Experimental configuration and simulation parameters}

In order to compare with the experimental observations of ASDEX Upgrade discharges, realistic plasma parameters and geometry are used in the modeling. As reference case, equilibrium reconstruction of the density, temperature and magnetic profiles are extracted from the discharge $\#31128$, where a strong ELM mitigation is observed in experiments \cite{Suttrop_PPCF17}. Note that the temperature and density profiles used in modeling are the experimental pre-ELM profiles before RMPs are applied, in order to study a peeling-ballooning unstable plasma. These profiles, extracted at the time $t=2.4s$ of the discharge $\#31128$, are plotted in Fig.\ref{fig:fig10}. The time evolution of the main plasma parameters during this discharge is plotted in Fig.2(a) in the reference \cite{Suttrop_PPCF17}. As in experiments, $n=2$ static magnetic perturbations are applied in modeling, with a nominal current of about $6$kAt flowing in the RMP coils. In the experiments, the modification of the differential phase $\Delta \Phi$ between the upper and lower coil currents allows to change the applied RMP spectrum and thus the plasma response to RMPs. In the discharge $\#31128$, a constant phase $\Delta \Phi = +90^{\circ}$ deg is steadily applied. During the discharge $\#30826$ operated with similar plasma profiles as in the discharge $\#31128$, the time variation of $\Delta \Phi$ from $+90^{\circ}$ to $-90^{\circ}$ showed that the strongest ELM mitigation is found for $\Delta \Phi = +90^{\circ}$ \cite{Kirk_NF15}. During this discharge, when $\Delta \Phi$ is reduced towards the opposite phase $-90^{\circ}$, a transient ELM-free phase is observed, suggesting that the stability limit previously reduced by effective RMPs is increased back to a threshold close to the stability limit without RMPs \cite{Chapman_PoP13}: hence, in this plasma configuration, RMPs have the smallest effect on ELM for  $\Delta \Phi \sim -90^{\circ}$.

The impact RMPs have on ELMs is modeled for these two opposite phases ($\Delta \Phi = +90^{\circ}$ and $-90^{\circ}$), for a nominal applied RMP amplitude (\textit{i.e.} the experimental amplitude) and for an amplitude increased by $50\%$. The modeling was done with the reduced MHD model of JOREK including the two-fluid diamagnetic rotation, the neoclassical poloidal friction and a source of toroidal rotation reproducing the experimental profile \cite{Orain_PoP13}. The parameters used are similar to the one described in Ref. \cite{Orain_NF17} and reproduce as accurately as possible the experimental situation. The main limitations of the modeling are the increased central resistivity $\eta_0 = 8.1 \times 10^{-7} \Omega.m$ (value one order of magnitude larger than the Spitzer value, but the $T_e^{-3/2}$ dependency is accounted for in the simulations, $T_e$ being the electron temperature) and the restriction of the simulations to the toroidal modes $n \leq 8$ to limit their computational costs. The impact of these limitations is discussed in section \ref{sec:discu}.

\subsection{General plasma response to RMPs} \label{subsec2}

Before presenting the interaction between ELMs and RMPs, the plasma response to RMPs (without ELMs), described in more details in \cite{Orain_NF17}, is summarized in order to make the following more understandable. In this paragraph, only the axisymmetric component ($n=0$) and the $n=2$ mode are included in the simulation, to prevent ELMs (generally constituted of modes $n \geq 4$ \cite{Mink_NF18}) from growing and to consider only the plasma response to RMPs. 

\begin{figure}[h!]
\centering
\includegraphics[width=\textwidth]{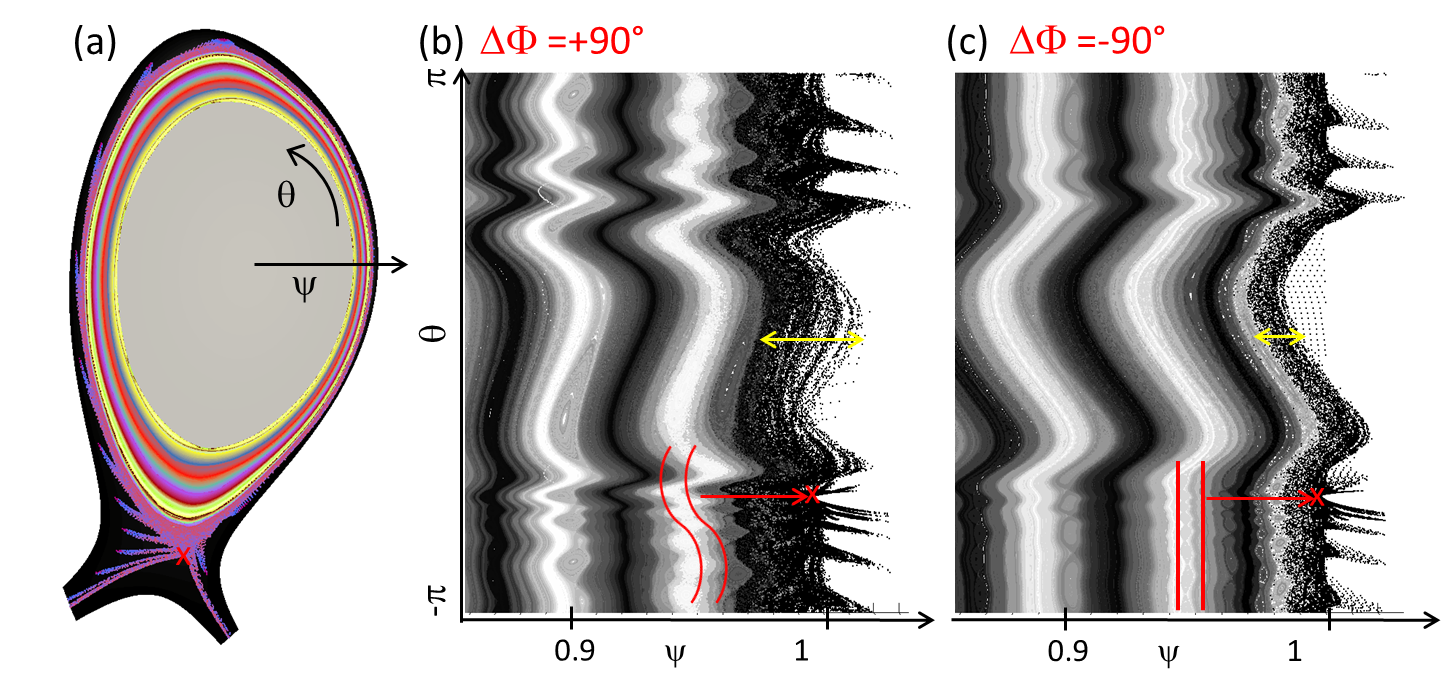}
\caption{Poincar\'e plot of the magnetic topology (a) in $[R,Z]$ coordinates (b-c) in radial and poloidal $[\psi,\theta]$ coordinates for the resonant ($\Delta \Phi = +90^{\circ}$) and non-resonant ($\Delta \Phi = -90^{\circ}$) configurations. (a) is given for the non-resonant case. The red crosses present the X-point location and the red lines sketch the strong kinking of the field lines near the X-point in the resonant ($\Delta \Phi = +90^{\circ}$) case, in opposition to the straight field lines observed near the X-point in the non-resonant ($\Delta \Phi = -90^{\circ}$) case. The yellow arrows highlight the radial width of the chaotic layer.}
\label{fig1}
\end{figure}

The $n=2$ RMPs are applied in the simulation and increased towards their nominal amplitude within $1000$ Alfv\'en times t$_{A}$ ($\sim 0.5$ ms), for different phases $\Delta \Phi$ between upper and lower coil currents. For all phases, the $3D$ equilibrium obtained presents similar main features: RMPs are screened by the flows in the plasma center, preventing magnetic islands from growing on central resonant surfaces. However, at the very edge, the large resistivity (related to the low temperature) allows island chains to form and a small chaotic layer is observed due to the overlapping of magnetic islands on the resonant surfaces $q \geq 9/4$ (for a normalized poloidal flux $\psi \geq 0.97$). Around the X-point, a typical lobe structure is induced by the destruction of the separatrix. The magnetic topology is plotted for the configuration $\Delta \Phi = -90^{\circ}$ in Fig. \ref{fig1}(a).

The main differences depending on the phase $\Delta \Phi$ actually appear only at the edge, as shown in Fig. \ref{fig1}(b-c). In the case when RMPs are strongly mitigated in experiments ($\Delta \Phi = +90^{\circ}$), the chaotic layer found in the modeling is largest, covering the edge plasma for $\psi \geq 0.97$, as indicated by the yellow arrow in Fig. \ref{fig1}(b). Near the X-point (highlighted by the red cross in Fig. \ref{fig1}(b)), the distortion of the field lines due to the amplification of peeling-kink modes, is also maximum in this configuration. Since both the ergodic layer width and the peeling-kink amplitude are maximum, we call it the ``resonant configuration". On the contrary, for the opposite phase $\Delta \Phi = -90^{\circ}$ when RMPs have least effect on ELMs in experiments, the smallest chaotic layer and the smallest distortion near the X-point are found in the modeling (Fig. \ref{fig1}(c)): we call it ``non-resonant configuration". 
As explained in Ref. \cite{Orain_NF17}, in the optimal (``resonant") configuration $\Delta \Phi = +90^{\circ}$, the poloidal coupling between the peeling-kink modes ($m>nq$) and the tearing modes (magnetic islands on $q=m/n$) allows for the penetration and amplification of the RMPs at the edge. In the opposite (``non-resonant") configuration ($\Delta \Phi = -90^{\circ}$), these modes remain at low amplitude because of the plasma screening.  

\subsection{RMP effect on ELMs}

From now on, all modes from $n=0$ to $8$ are included in the modeling to study the effect of ``penetrated" RMPs on ELMs in the two configurations described in the previous paragraph (section \ref{subsec2}). 
As a comparison, a simulation is also run without RMPs for the same plasma parameters. 
In a first step, in order to particularly highlight the saturation of the modes in the simulations, the mode coupling was enhanced  by increasing the anomalous viscosity (the dynamic viscosity is set to $3.7 \times 10^{-7} kg/(m.s)$ corresponding to a kinematic viscosity of $2m^2/s$): therefore the anomalous viscous term dominates over the neoclassical friction term, resulting in an overestimated stabilization of the modes. In the next section \ref{sec:sec3}, the reduction of the anomalous viscosity by a factor of $10$ allows the perpendicular dissipation to be mainly ruled by the neoclassical friction and to reproduce more realistically the threshold between ELM mitigation and ELM suppression.

\begin{figure}[h!]
\centering
\includegraphics[width=\textwidth]{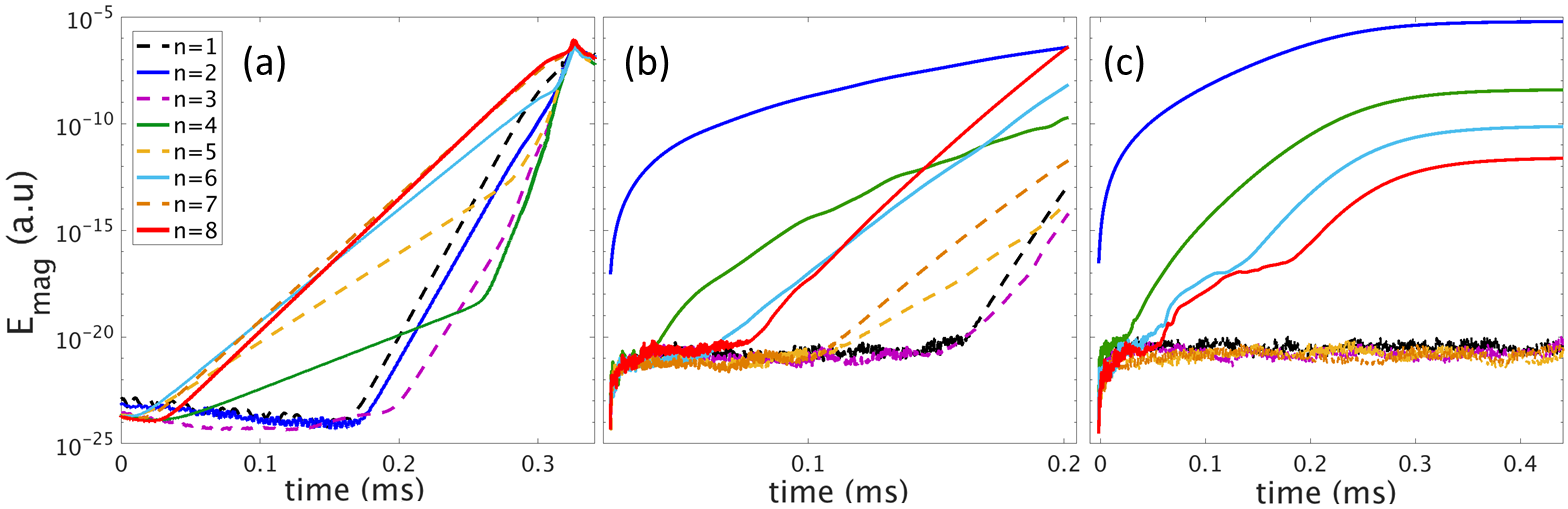}
\caption{Time evolution of the magnetic energy of the toroidal modes $n=1-8$ in the following cases: (a) ELM without RMP. (b) ELM with non-resonant RMPs. (c) ELM with resonant RMPs. Units are arbitrary but the normalization is the same in all the figures of the paper.}
\label{fig2}
\end{figure}

In the simulation without RMPs, since the experimental profiles before RMP application are used as input for modeling, the plasma is unstable. As shown in Fig. \ref{fig2}(a), peeling-ballooning modes with dominant mode numbers $n=6, 7$ and $8$ grow exponentially. Their non-linear coupling generate the growth of lower $n$ modes\cite{Krebs}, until all modes saturate and induce the ELM crash. Note that linear simulations show that $n \geq 9$ modes are also linearly unstable, with a growth rate smaller than $n=7$ and $8$. Since these two most unstable modes dominate the linear growth and the low $n \le 6$ modes are non-linearly dominant \cite{Mink_NF18}, it was chosen to discard the $n \geq 9$ modes to reduce the computation time of the simulations.

This ``natural" ELM without RMPs is compared to the cases with RMPs. Fig. \ref{fig2}(b) and (c) present respectively the time evolution of the magnetic energy of the different modes when ``non-resonant" ($\Delta \Phi = -90^{\circ}$) and ``resonant" ($\Delta \Phi = +90^{\circ}$) RMPs are applied. In both cases, we observe the growth of the $n=2$ mode induced by the RMP application. In non-resonant configuration (Fig. \ref{fig2}(b)), the $n=4$ mode growth follows the dynamics of $n=2$, due to quadratic coupling (cross-terms $[n=2] \times [n=2]$). However, the $n=6$ and $8$ modes grow exponentially with a growth rate close to the one without RMP, leading to an ELM crash of similar amplitude as in the case without RMPs. Note that the odd modes present a growth delayed by the RMP application but they finally develop exponentially.  
On the contrary, when resonant RMPs are applied (Fig. \ref{fig2}(c)), the $n=2$ mode driven by RMPs has a larger amplitude, due to the stronger penetration. This amplitude is large enough to allow the toroidal coupling of all even modes with the $n=2$ mode: these modes all follow the same dynamics driven by RMPs and saturate at an amplitude lower than the saturation level of the ELM without RMPs. This saturation induces the mitigation or the suppression of the ELMs. It is interesting to observe that the $n=2$ symmetry imposed by RMPs also prevents the growth of odd modes in resonant configuration. The mechanism of the ELM stabilization by non-linear mode coupling, observed here, was first described for a JET case in Ref. \cite{Becoulet_PRL14}.
 
%%% visco dyn =10-6 = 3.7e-7 kg/(m.s) -> visco kin = 2 m2/s
%%% Nouveaux cas: visco = 10-7 = 3.7e-8 kg/(m.s) -> visco kin = 0.2 m2/s
%%% eta = 3.10-7 -> eta_SI = 8.1.10-7 Ohm.m
\begin{figure}[h!]
\centering
\includegraphics[width=0.6\columnwidth]{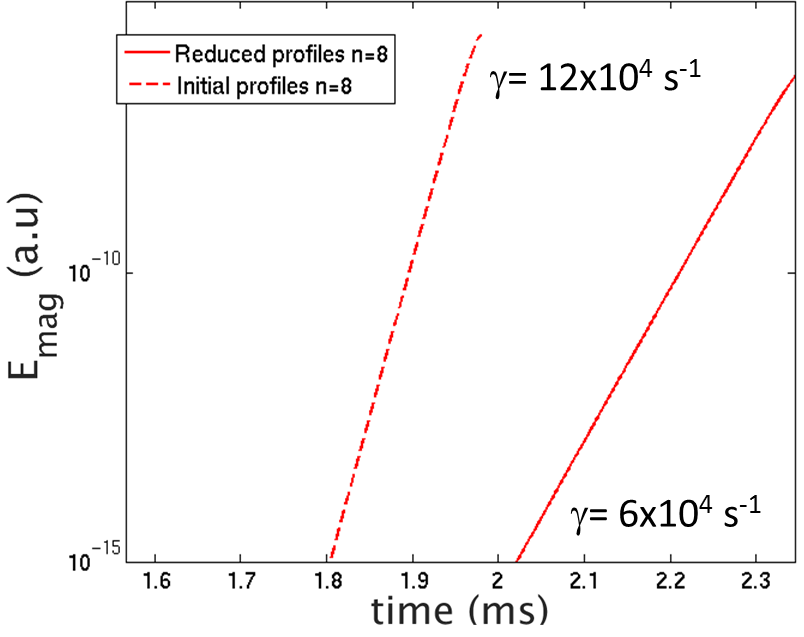}
\caption{Magnetic energy of the $n=8$ mode without RMP for the initial pressure profile (as in Fig. \ref{fig2}(a), dashed line) and for a reduced pressure profile (as in Fig. \ref{fig2}(c), full line). The corresponding growth rate $\gamma$ is given.}
\label{fig:fig3}
\end{figure}
In the resonant case when ELM suppression is observed in modeling (Fig. \ref{fig2}(c)), RMPs induce an enhanced perpendicular transport due to the formation of a chaotic layer at the edge, therefore the edge pressure gradient is reduced. In order to check if the ELM suppression is due to the reduction of the pressure gradient under the peeling-ballooning stability limit, the following test is done: a simulation is performed with a pressure gradient reduced to the same level as in the ELM suppression case, but without applying RMPs. The modified current density profile is also considered. In this simulation, the exponential growth of the $n=8$ Edge Localized mode is observed, leading to an ELM crash. As shown in Fig. \ref{fig:fig3}, the growth rate is reduced by a factor of two as compared to the natural ELM with unchanged pressure profile, but the ELM is still unstable and leads to a crash. It shows that the ELM suppression is indeed caused by the ELM saturation induced by the mode coupling, rather than by the reduction of the pressure gradient below the ELM stability limit. 
%This observation is still valid for the ELM suppression case described in the following section \ref{sec:sec3}: even with the reduced pressure gradient induced by the radial transport during ELM suppression, the plasma remains peeling-ballooning unstable, and the mode coupling is necessary to explain the ELM suppression.

\section{Threshold between ELM mitigation and suppression} \label{sec:sec3}

\subsection{Bifurcation from mitigation to suppression} \label{sec:sec31}
\begin{figure}[h!]
\centering
\includegraphics[width=0.6\columnwidth]{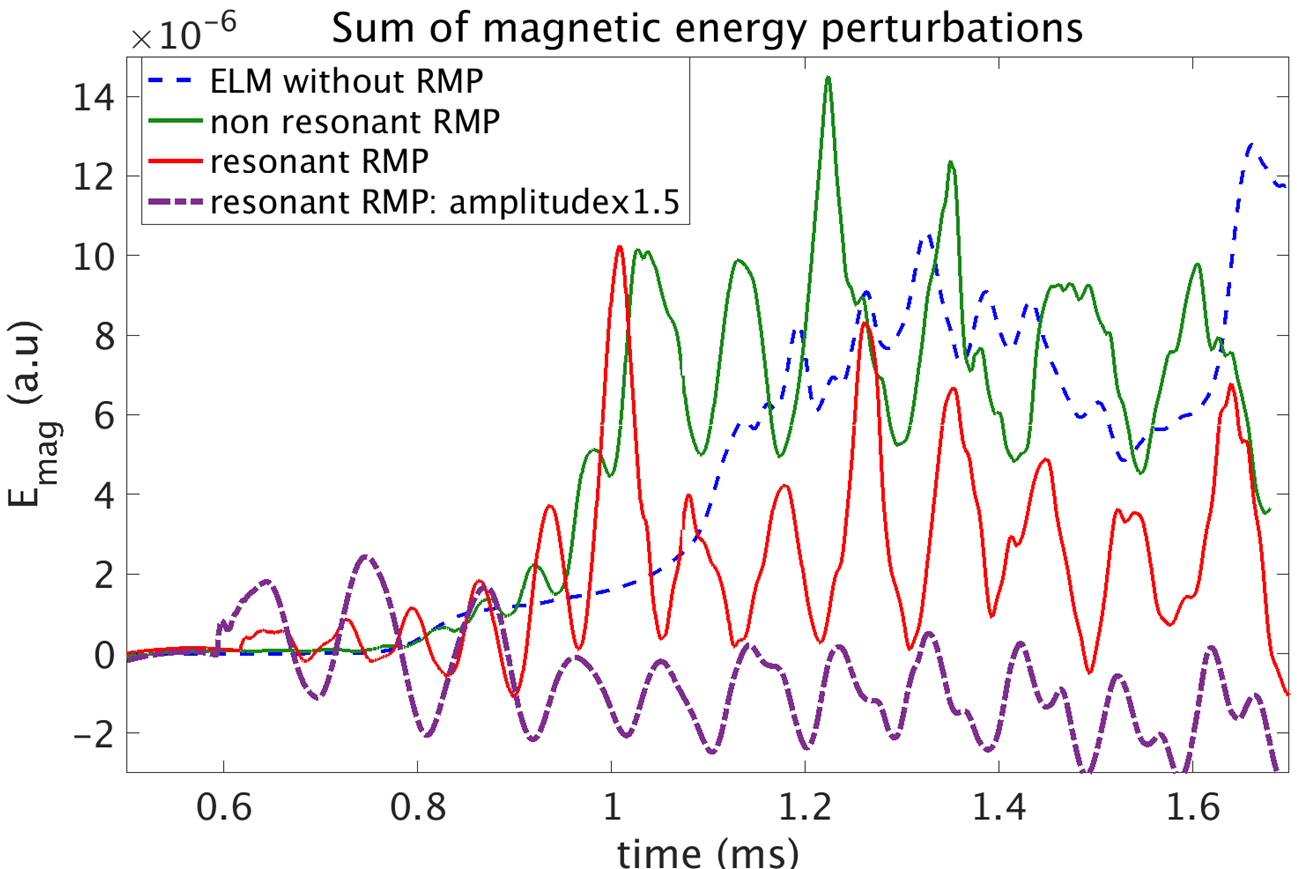}
\caption{The sum of the magnetic energies associated to the non-axisymmetric components is plotted in the cases of an ELM without RMP (dash blue), with non-resonant RMP (green), with resonant RMP at nominal amplitude (red), with resonant RMP at amplitude increased by $50\%$ (dash-dotted line in purple). The value corresponding to the magnetic energy perturbation of the RMPs alone is subtracted. In the non-linear evolution, this difference can also drop slightly below zero.}
\label{fig:fig4}
\end{figure}

In this section, the reduced viscosity allows to reveal the fine mechanisms distinguishing ELM mitigation from  ELM suppression. In order to characterize the ELM energy, the energy of the modes can be separated into two parts: the externally injected static energy induced by RMPs and the intrinsic (non-static) energy corresponding to the energy of the ELM, mitigated or not by RMPs. The sum of the intrinsic magnetic energy of all perturbations is plotted in  
Fig. \ref{fig:fig4} for different cases: natural ELM without RMP, ELM with non-resonant RMPs applied ($\Delta \Phi = -90^{\circ}$) at nominal amplitude and ELM with resonant RMPs applied ($\Delta \Phi = +90^{\circ}$) at nominal and increased amplitude (by $50\%$). 
We observe that when non-resonant RMPs are applied at nominal amplitude (green line), the ELM reaches the same level of energy as for the uncontrolled ELM (blue dashed line), showing the inefficiency of non-resonant RMPs. However, when resonant RMPs are applied at nominal amplitude (red line), the ELM energy is reduced significantly (red line), as in ELM mitigation observed in experiments. Furthermore, when the resonant RMP amplitude is increased by $50\%$ (purple line), the intrinsic energy vanishes, corresponding to full ELM suppression. It means that above a certain level of energy injected by the RMP penetration, the toroidal coupling of the edge modes (normally unstable) with the $n=2$ static mode induced by RMPs is large enough to force them to saturate.
Note that when non-resonant RMPs are applied at a larger amplitude multiplied by 2, ELM suppression is still not observed, which confirms the importance of the resonant condition to obtain ELM suppression.

\begin{figure}[h!]
\centering
\includegraphics[width=\textwidth]{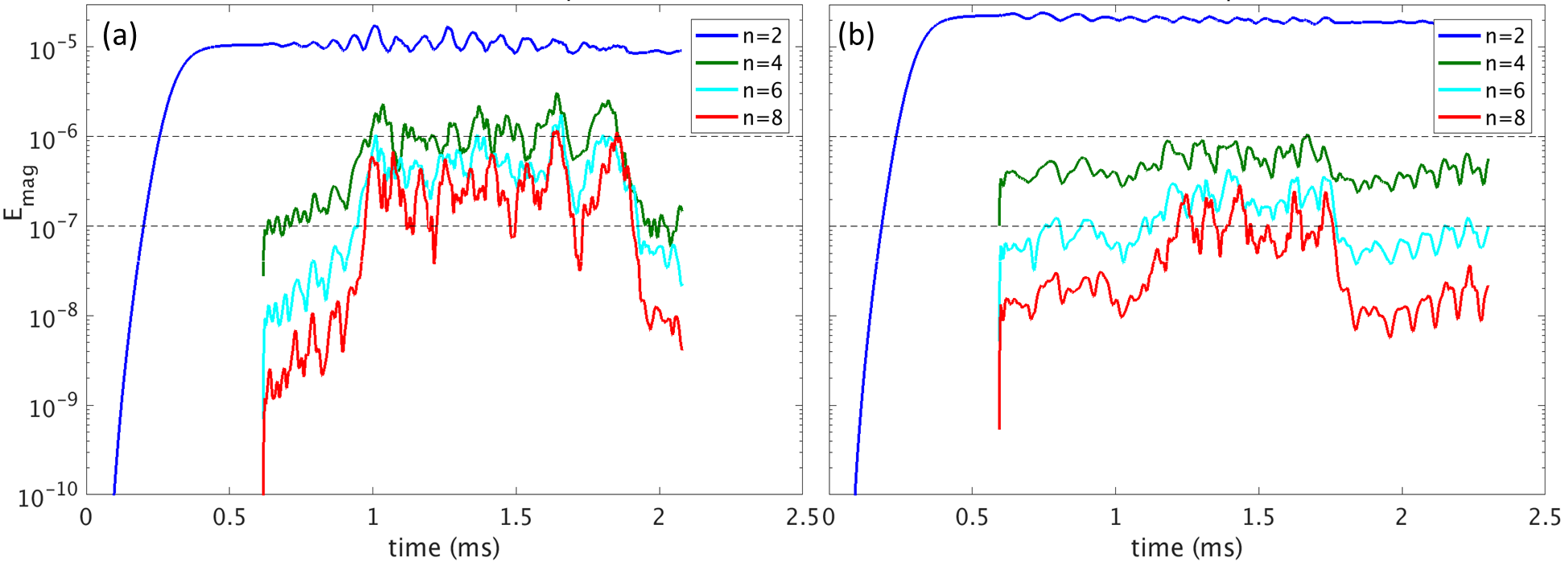}
\caption{Magnetic energy of the modes $n=2,4,6$ and $8$ in ELM mitigation case obtained for nominal RMP amplitude (a) and in ELM suppression case obtained at RMP amplitude increased by $50\%$ (b).}
\label{fig:fig6}
\end{figure}

The details of the energy evolution in ELM mitigation (for nominal resonant RMP amplitude) and ELM suppression regimes (for increased RMP amplitude) is provided in Fig. \ref{fig:fig6} (respectively (a) and (b)). The simulation is started with the $n=2$ mode alone, when RMPs are applied at $t \approx 0.1$ms and the other even modes $n=4, 6$ and $8$ are added at $t \approx 0.6$ms. Note that the odd modes were discarded from the simulation in order to reduce the computational time, since they anyway remain at the noise level when ``resonant" RMPs are applied at sufficient amplitude, as shown in the previous section.

 In the ELM mitigation case (a), when the modes $n=4, 6$ and $8$ are included in the simulation, an initial ``linear" phase is observed (from $t \approx 0.6$ms to $1$ms). All along this phase, these modes grow exponentially, with a reduced growth rate as compared to the ELM without RMPs. Then, in the non-linear phase (from $t \approx 1$ms), the modes reach a saturation amplitude below the amplitude of the ELM without RMPs. During this ELM-mitigated regime, which lasts until $t \approx 1.9$ms, the $n=4, 6$ and $8$ modes are rotating in the $E \times B$ or electron diamagnetic direction and plasma filaments are expelled in the opposite (ion diamagnetic) direction: these modes behave as peeling-ballooning modes in a similar way as in the natural ELM regime, but with a much smaller amplitude and hence they induce a much smaller exhaust. Note that the interaction between the rotating unstable modes and the $n=2$ perturbation induced by RMPs forces the $n=2$ mode to co-rotate with the other modes, at the frequency $f \sim m V_\theta / (2 \pi r_{res}) \sim n q V_\theta / (2 \pi r_{res}) \sim 10-25Hz$. $q, r_{res}$ and $V_\theta$ are respectively the safety factor, minor radius and poloidal rotation at the resonant surface where the amplitude is maximal (for $q=4$); $n=2$ and $m=nq$ are respectively the toroidal and poloidal mode numbers. During ELM mitigation, the energy loss rate (respectively the particle loss rate) between $t = 1.1$ms and $1.3$ms is reduced by one third (respectively by $40\%$) as compared to the ``natural" ELM relaxation.

In comparison, in the ELM suppression case (Fig. \ref{fig:fig6}(b)), the amplitude of the static $n=2$ mode is larger, since a larger current is applied in RMP coils ($+50\%$) and thus RMPs penetrate more significantly. During the initial phase when the modes $n>2$ are added in the simulation at $t \approx 0.6$ms, the modes $n=4, 6$ and $8$ start with an initial amplitude imposed by the coupling with $n=2$ and remain at the same level, until they grow and saturate at an even smaller amplitude than in the ELM mitigation case (around one order of magnitude smaller). In the ELM suppression regime, the energy and particle loss rates are strongly reduced as compared to the ``natural" ELM relaxation, by $60\%$ and $80\%$ respectively. The remaining radial transport can be explained by the chaos induced by the RMP penetration, as discussed in \ref{sec:chaos}.
  
\begin{figure}[h!]
\centering
\includegraphics[width=\columnwidth]{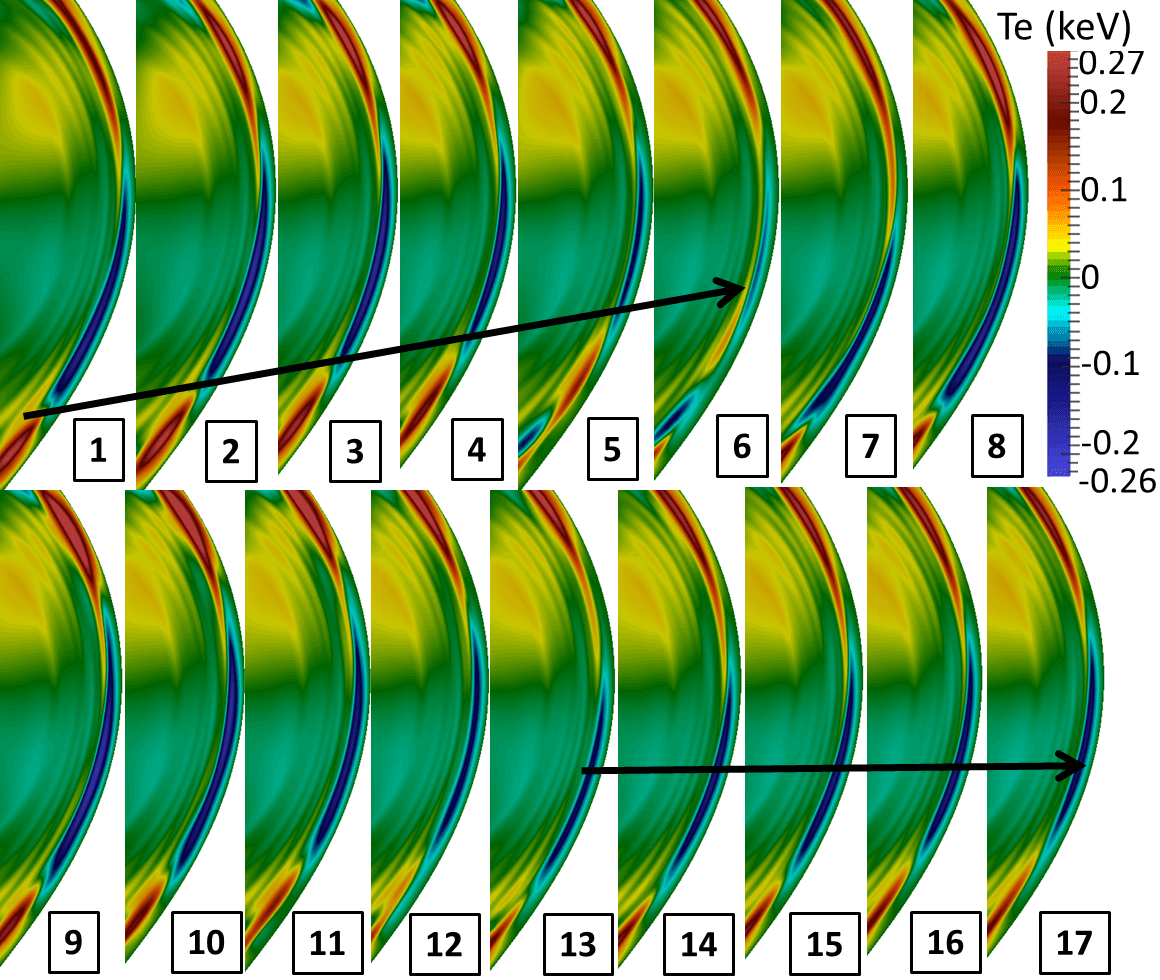}
\caption{Perturbation of the electron temperature at the outboard midplane in the ELM suppression regime for 17 time steps between $t=1.652$ms and $1.802$ms. A constant time $\Delta t=8.8 \mu s$ separates two consecutive steps. The arrows highlight the mode rotation at the beginning (pictures $1$ to $6$, from $t=1.652$ms to $1.70$ms), followed by a sustained static perturbation.}
\label{Screenshots_suppr}
\end{figure}

Still in the ELM suppression case (Fig. \ref{fig:fig6}(b)), the edge modes $n=4, 6$ and $8$ are initially rotating (as in the ELM mitigation case) and also force the $n=2$ mode to co-rotate with them, but the modes suddenly start to slow down around $t \approx 1.6$ms, until they stop rotating and remain static around $t \approx 1.75$ms. The perturbation of the electron temperature at the outboard midplane is plotted in Fig. \ref{Screenshots_suppr} for $17$ time steps between $t=1.652$ms and $1.802$ms. It highlights the propagation of the last rotating perturbation (pictures $1$ to $6$, from $t=1.652$ms to $1.70$ms) followed by a sustained static perturbation. The mechanism of the mode braking and the possible correlation between ELM suppression and mode braking are  described in the next subsection \ref{sec:rotation}.  

\begin{figure}[h!]
\centering
\includegraphics[width=\columnwidth]{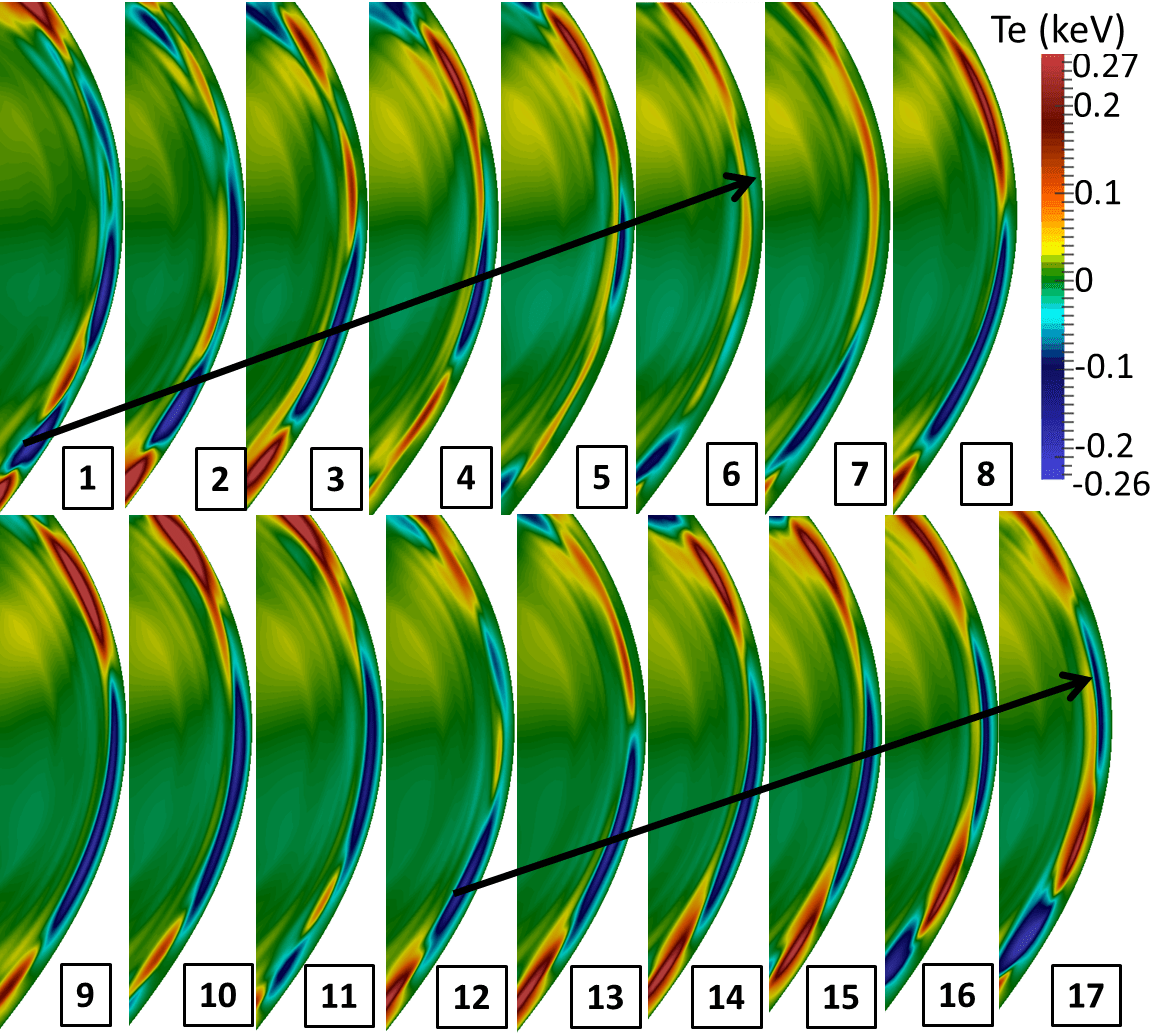}
\caption{Perturbation of the electron temperature at the outboard midplane in the ELM mitigation regime, plotted for the same time steps as in the ELM suppression regime (Fig. \ref{Screenshots_suppr}). The arrows underline the mode rotation in the $E \times B$ or electron diamagnetic direction.}
\label{Screenshots_mitig}
\end{figure}

In comparison, Fig. \ref{Screenshots_mitig} shows that for the same time slices, the edge modes keep on rotating in the ELM mitigation regime. Note that the modes rotate counterclockwise (\textit{i.e.} in the $E \times B$ direction) with a larger speed than in the early phase of the rotation braking observed in ELM suppression, as shown by the black arrows following the perturbation in time. The qualitative comparison of the rotation with experimental measurements is provided in the next subsection \ref{sec:rotation}.

Furthermore, it is interesting to notice that in the ELM mitigation regime, all the even modes (except $n=2$) reach the same order of magnitude in the non-linear phase. Indeed, when the perturbation of the magnetic energy is maximal (for $t=1.64$ms), the ratio between the amplitudes of the energies of the different modes is the following: $E_{n=4}/E_{n=6}=2.5$ and $E_{n=6}/E_{n=8}=1.05$. The amplitude of these modes is also non-negligible as compared to the $n=2$ mode amplitude: $E_{n=2}/E_{n=4}=3.8$. On the contrary, in the ELM suppression regime, the $n=2$ mode number clearly dominates over the other modes (at the maximal energy perturbation for $t=1.67$ms, $E_{n=2}/E_{n=4}=18.9$) and the next mode number $n=4$ is also dominant over the larger mode numbers: $E_{n=4}/E_{n=6}=4.5$ and $E_{n=6}/E_{n=8}=2.8$. During the natural ELM without RMP, the medium modes $n \ge 6$ dominate over the lower modes. When RMPs are applied, the different energy redistribution during ELM mitigation and ELM suppression therefore shows that the application of the $n=2$ RMPs had imposed the modes $n=4, 6$ and $8$ to remain at a lower level in ELM suppression as compared to ELM mitigation. In other words, a stronger energy transfer has been operated from medium toroidal $n$ mode numbers towards lower $n$ modes (principally towards $n=2$).

\subsection{Mechanism of the mode braking during ELM suppression} \label{sec:rotation}
\begin{figure}[h!]
\centering
\includegraphics[width=\textwidth]{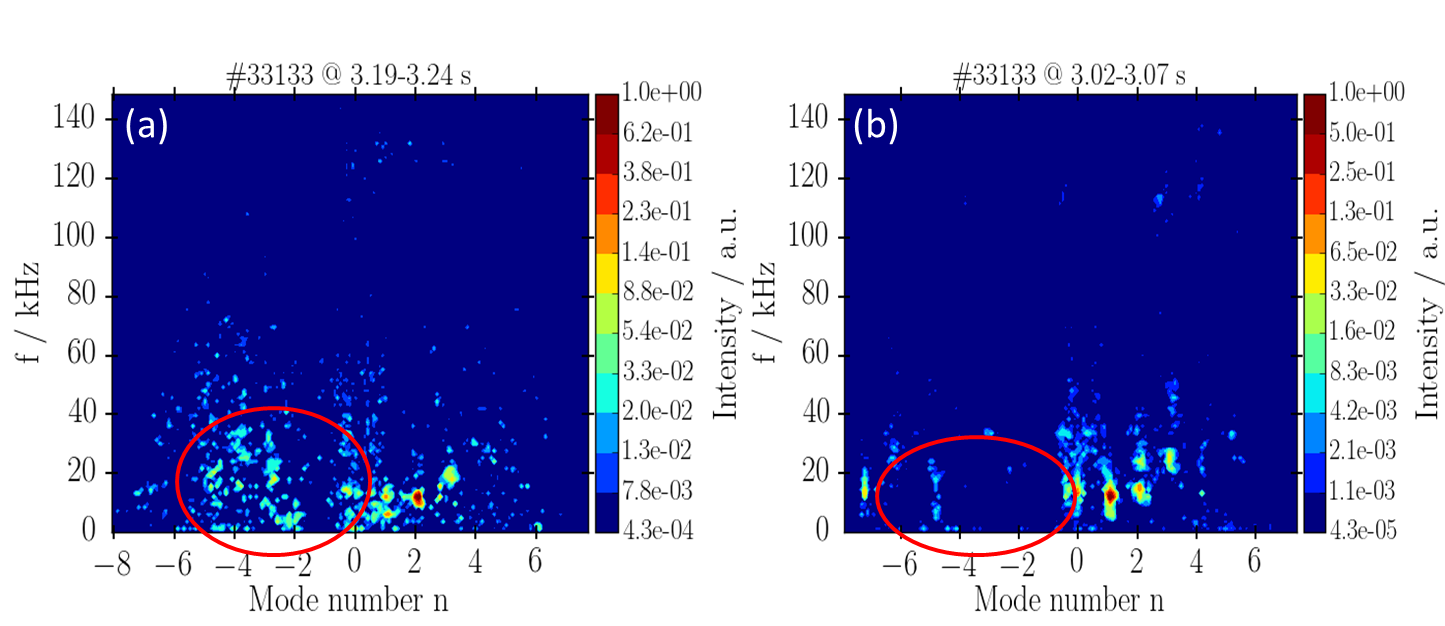}
\caption{Experimental mode spectrum in ELM mitigation (a) and ELM suppression (b) phases of the discharge $\#33133$: the intensity of the modes is plotted depending on their frequency and toroidal mode number.}
\label{fig:fig5}
\end{figure}

The rotation of the different modes can be directly compared with the experimental observations. In the discharge $\#33133$, the bifurcation from ELM mitigation to ELM suppression is observed \cite{Suttrop_NF18}. The toroidal mode spectrum is calculated from the measurements of the magnetic fluctuations, which means that only fluctuating modes can be observed in the spectrograms. In Fig. \ref{fig:fig5}, negative $n$ mode numbers correspond to modes rotating in the electron diamagnetic direction and indicates the movement of edge modes, while positive mode numbers describe core modes rotating in the ion diamagnetic direction \cite{Mink_NF18}. The comparison of the ELM mitigation (Fig. \ref{fig:fig5}(a)) and ELM suppression (Fig. \ref{fig:fig5}(b)) regimes shows that rotating modes are observed at the edge in the ELM mitigation phase. On the contrary, in the ELM suppression phase, almost no fluctuating modes are present, which means that the ELMy regime characterized by rotating edge-localized modes is replaced by a regime with either static modes or no mode at all. These observations corroborate the idea that ELM suppression regime might consist of static saturated modes instead of rotating modes in natural ELMs or mitigated ELMs, as found in modeling.

\begin{figure}[h!]
\centering
\includegraphics[width=0.6\columnwidth]{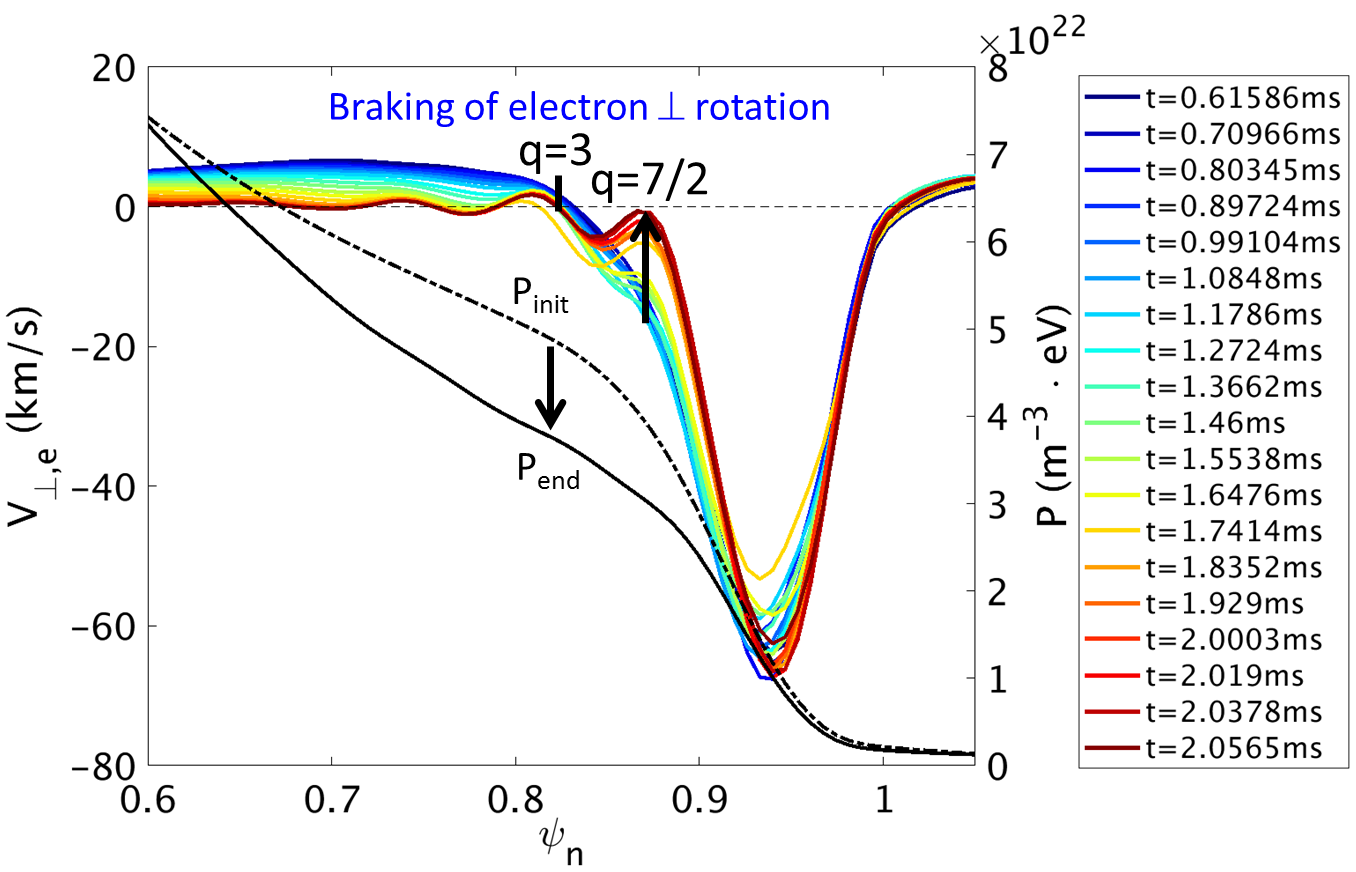}
\caption{Radial profile of the perpendicular electron rotation evolving in time between $t=0.6ms$ (blue) and $t=2.05ms$ (red) in the ELM suppression case. The resonant braking on the $q=7/2$ surface at the pedestal top is highlighted by a black arrow. The radial pressure profile  at the times $t=0.6ms$ (dash-dotted line, black) and $t=2.05ms$ (full line, black) is overlaid in order to locate the pedestal top and the strong gradient region.}
\label{fig:fig7}
\end{figure}

In the modeling of the ELM suppression case described above (subsection \ref{sec:sec31}), the braking of the edge modes until they become static seems to be induced by the braking of the electron perpendicular rotation. Theorized first in Ref. \cite{Fitzpatrick} and refined for RMPs in Ref. \cite{Nardon_NF10}, the bifurcation from a screened configuration towards the penetration of RMPs is correlated with the sudden braking of the electron perpendicular rotation on the resonant surfaces induced by electromagnetic torque. In the simulation of the ELM suppression presented in Fig. \ref{fig:fig6}(b), such a bifurcation is observed at the pedestal top, as plotted in Fig. \ref{fig:fig7}. On the resonant surfaces $q=3$ (for a normalized poloidal magnetic flux $\psi=0.82)$, the perpendicular electron velocity is already close to zero at the beginning, allowing the penetration of magnetic islands. On the $q=7/2$ surface (for  $\psi=0.88)$, located near the pedestal top, the abrupt reduction of the perpendicular electron flow occurs between $t=1.6$ and $2$ms. This rotation braking likely induces the braking of the modes: edge modes are forced to be in phase with the penetrated mode induced by RMPs and thus to become static. 
The mode penetration inducing the sudden braking of the electron perpendicular velocity was observed in ELM suppression regimes in the experiments of DIII-D \cite{Wade_NF15, Nazikian_PRL15} and KSTAR \citep{Lee_PRL16, Lee_IAEA18}. The reduction below $1km/s$ of the velocity of saturated edge modes observed in non-bursting ELM suppression regimes of KSTAR \citep{Lee_PRL16, Lee_IAEA18}, as well as the phase-locking of edge peeling-kink modes with static RMPs in grassy-ELM suppression regimes of DIII-D \cite{Nazikian_NF18} is consistent with the edge mode locking found in this modeling. In ASDEX Upgrade, the zero-crossing of the electron perpendicular flow at the pedestal top is observed in some ELM suppression discharges, but not all of them \cite{Suttrop_NF18}, meaning that the RMP-induced mode penetration at the pedestal top may be one of the ELM suppression mechanisms, but other mechanisms may exist too. 

\subsection{Impact of the electric field evolution} \label{sec:elec}

\begin{figure}[h!]
\centering
\includegraphics[width=0.6\columnwidth]{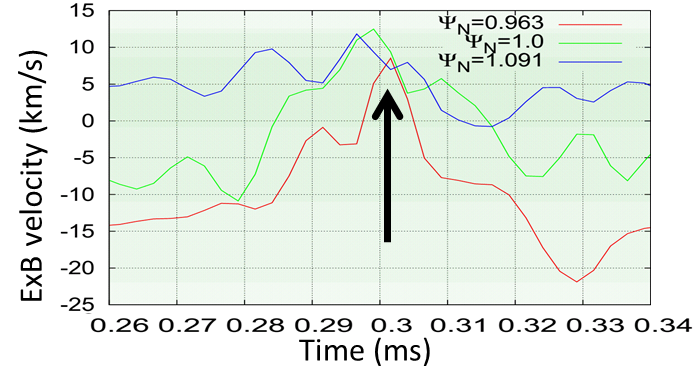}
\caption{Time evolution of the $E \times B$ velocity in an ELM simulation (without RMP) at different radial positions: just inside the separatrix ($\psi=0.963$, red), at the separatrix ($\psi=1$, green) and just outside the separatrix ($\psi=1.091$, blue). The black arrow shows the moment when filaments are expelled out of the separatrix.}
\label{fig:fig8}
\end{figure}

It is interesting to notice that the expulsion of filaments through the separatrix can be characterized by the evolution of the radial electric field. Indeed, in the modeling, at the time when filaments are expelled from the plasma edge during the ELM crash, the $E \times B$ velocity and thus the radial electric field $E_r$ in the pedestal evolves from a large negative value to a zero or even positive value at the very edge. The plot of the $E \times B$ rotation close to the separatrix during the simulation of an ELM crash (Fig. \ref{fig:fig8}, for the discharge $\#23221$ described further in \cite{Orain_EPS16}) shows the transient reversal of the $E \times B$ velocity at the very edge ($\psi=0.963$) when the ELM filaments are expelled. The vanishing of the minimum $E_r$ during an ELM has also been observed in experiments using charge exchange recombination spectroscopy measurements: during the ELM crash, the edge $E_r$ collapses to very small $L$-mode like values and then recovers again to the pronounced $E_r$ well about $4$ms after the ELM crash \citep{Wade_PoP2005, Viezzer_NF13, Cavedon_PPCF2017}. The evolution of the $E_r$ profile is due to the fact that the non-linear Maxwell stress induces a strong shear of the plasma filaments close to the separatrix \cite{Huysmans_NF07}, which can also allow cold pulses to penetrate inside the separatrix \cite{Trier_PPCF19}. 

\begin{figure}[h!]
\centering
\includegraphics[width=\textwidth]{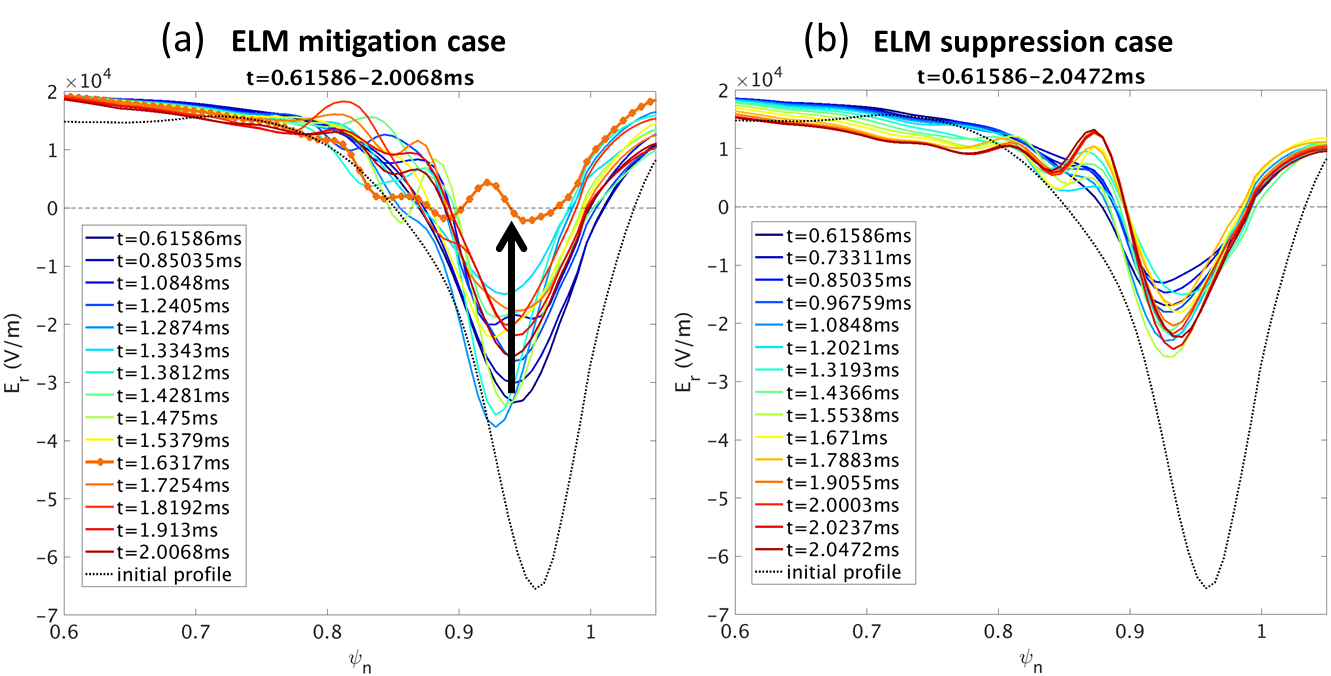}
\caption{Radial profiles of the radial electric field in the cases of ELM mitigation (a) and ELM suppression (b), for different times between $0.6$ and $2.05ms$. In the ELM mitigation case, $E_r$ vanishes at the time when the magnetic activity of all modes is maximal ($t=1.6ms$). The initial $E_r$ profile is also plotted in back dotted line in order to show the reduction of the $E_r$ well induced by the RMP application.}
\label{fig:fig9}
\end{figure}

For the discharge considered here ($\#31128$, \ref{sec:sec31}), the natural ELM crash (without RMPs) is also characterized by the transient vanishing of the $E \times B$ velocity and the radial electric field $E_r$ at the edge when the filaments are expelled. Interestingly, in the ELM mitigation case, the plot of $E_r$ (Fig. \ref{fig:fig9}(a)) shows that $E_r$ vanishes  in the pedestal at the moment when the magnetic activity of the edge modes is maximal (for $t=1.6ms$). This is in line with the fact that plasma filaments are crossing the separatrix at this time.  

On the contrary, in the ELM suppression regime, the radial electric field does not vanish in the pedestal (Fig. \ref{fig:fig9}(b)), reflecting the absence of filament exhaust in ELM suppression. 

Another feature displayed in Fig. \ref{fig:fig9}(a-b) is the fact that the radial electric field in the pedestal is smaller in absolute value in the ELM suppression regime ($-10$ to $-20kV/m$) than during ELM mitigation ($-20$ to $-35kV/m$), because of the larger amplitude of the penetrated RMPs. The initial electric field profile before RMP application is also plotted in Fig.  \ref{fig:fig9}(a-b): it shows that compared to the pre-RMP application, the $E_r$ well was strongly reduced during ELM mitigation (factor 2) and even more during ELM suppression (factor $3-4$). Turbulence cannot be observed in this modeling since turbulent transport is represented for simplicity by diffusive transport terms. Nevertheless, the reduction of the $E_r$ well (and thus also of the $E_r$ shear) observed in this MHD modeling may have a link with the increased broadband turbulence observed in ELM suppression states in ASDEX Upgrade experiments \cite{Leuthold_EPS18}, as suggested by turbulence modeling \citep{Kim_IAEA18, Kim_NF19}, where enhanced fluctuations are induced by energy transfer from kinetic energy ($E \times B$ flow) to magnetic energy (magnetic fluctuations). 

\begin{figure}[h!]
\centering
\includegraphics[width=\textwidth]{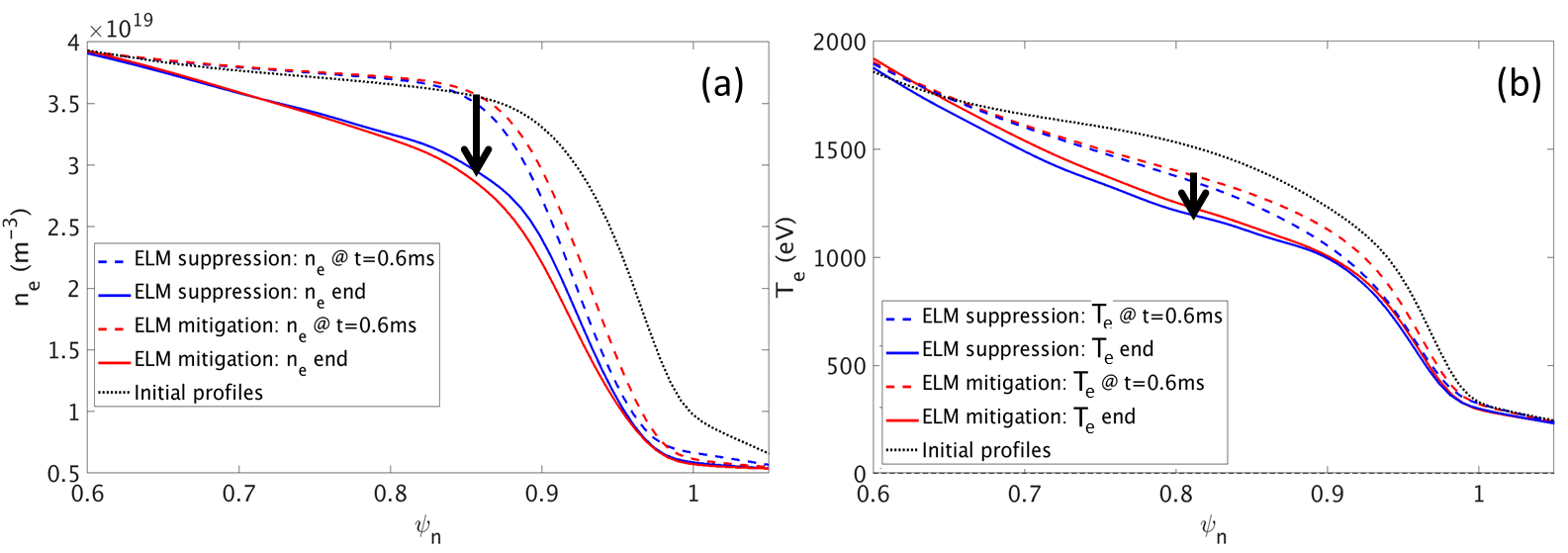}
\caption{Radial profiles of electron density (a) and temperature (b) at the starting time $t=0ms$ of the simulations (profiles before RMP application, black dotted line) and in the cases of ELM mitigation (red) and ELM suppression (blue) by RMPs, from the time when $n>2$ modes are included in simulation ($t \approx 0.6ms$, dash) to the time after the mode activity ($t \approx 2.05ms$, full line).}
\label{fig:fig10}
\end{figure}

Furthermore, the evolution of the electron temperature and density profiles (Fig. \ref{fig:fig10}) shows a very similar degradation of the pedestal density and temperature in both the ELM mitigation and the ELM suppression cases. A more realistic transport model might also be necessary to reproduce more accurately the large density pumpout observed in experiments \cite{Suttrop_PPCF17, Suttrop_NF18}. However the similar pressure reduction found for both regimes in the modeling highlights again the fact that the reduction of the pressure gradient cannot explain alone the ELM suppression: a strong mode-coupling is needed to saturate the edge modes at a low level and to induce the edge-localized mode braking.
  
\subsection{Evolution of the stochasticity level} \label{sec:chaos}
\begin{figure}[h!]
\centering
\includegraphics[width=0.6\columnwidth]{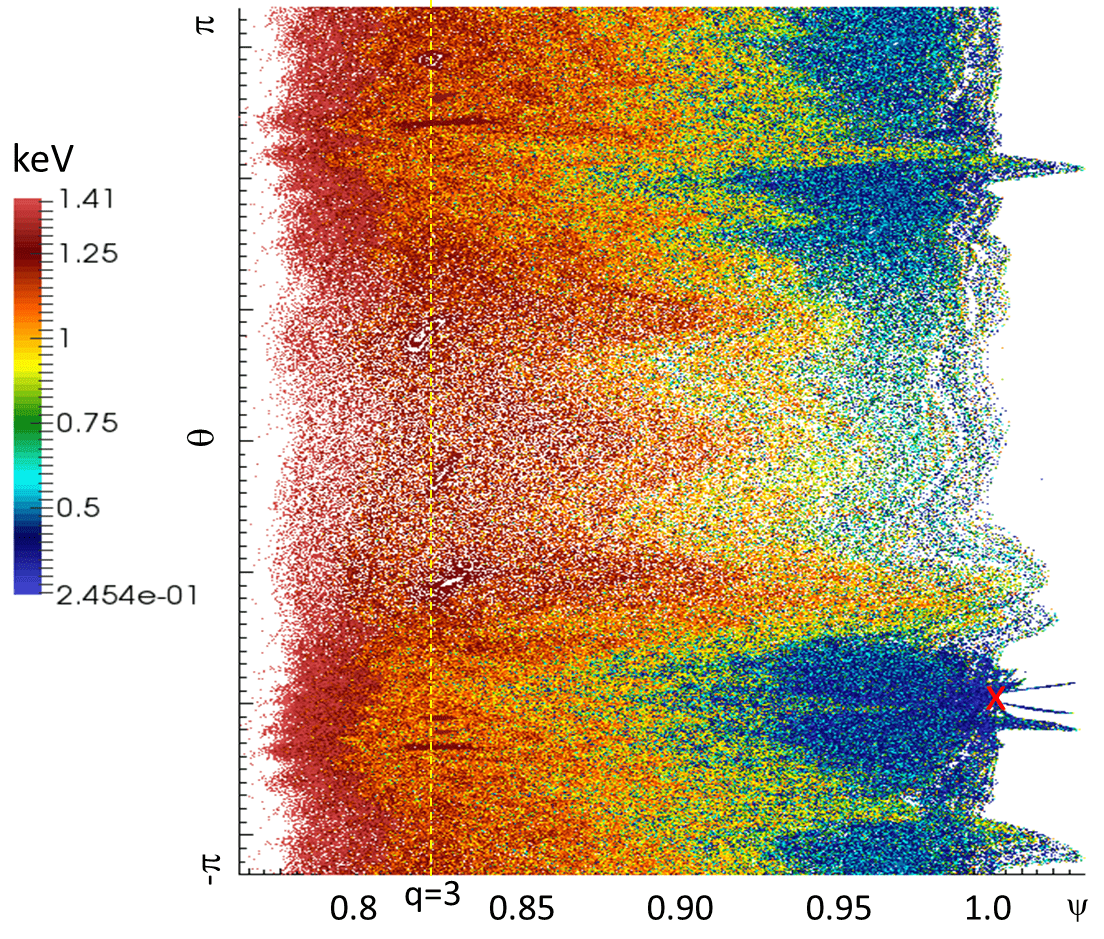}
\caption{Poincar\'e plot of the magnetic topology for the ELM without RMP. The field lines are followed for 100 toroidal turns and their end point is coloured according to the temperature of the starting point. The x- and y- axis are the radial $\psi$ and poloidal $\theta$ coordinates. The $q=3$ surface is highlighted by the yellow dotted line and the red cross shows the X-point location.}
\label{fig:stochasticity}
\end{figure}

Another important parameter characterizing the dynamics of the ELMs -- controlled or not -- is the level of chaos or stochasticity of the magnetic field lines. In the modeling of the crash of a natural ELM without RMPs, a wide stochastization of the magnetic field at the plasma edge is observed from the magnetic surface characterized by $q=3$ to the edge \cite{Trier_PPCF19, Hoelzl_CPP18}. In the ELM considered here, the full stochastization from the $q=3$ surface (located around $\psi =0.82$) is observed on Fig. \ref{fig:stochasticity}. In this Poincar\'e plot, field lines are followed for 100 toroidal turns and their end point is coloured according to the temperature of the starting point. It shows that long radial patterns extend from $\psi = 0.8$ to the edge, due to the enhanced transport following the broken field lines in the stochastic region.

\begin{figure}[h!]
\centering
\includegraphics[width=\textwidth]{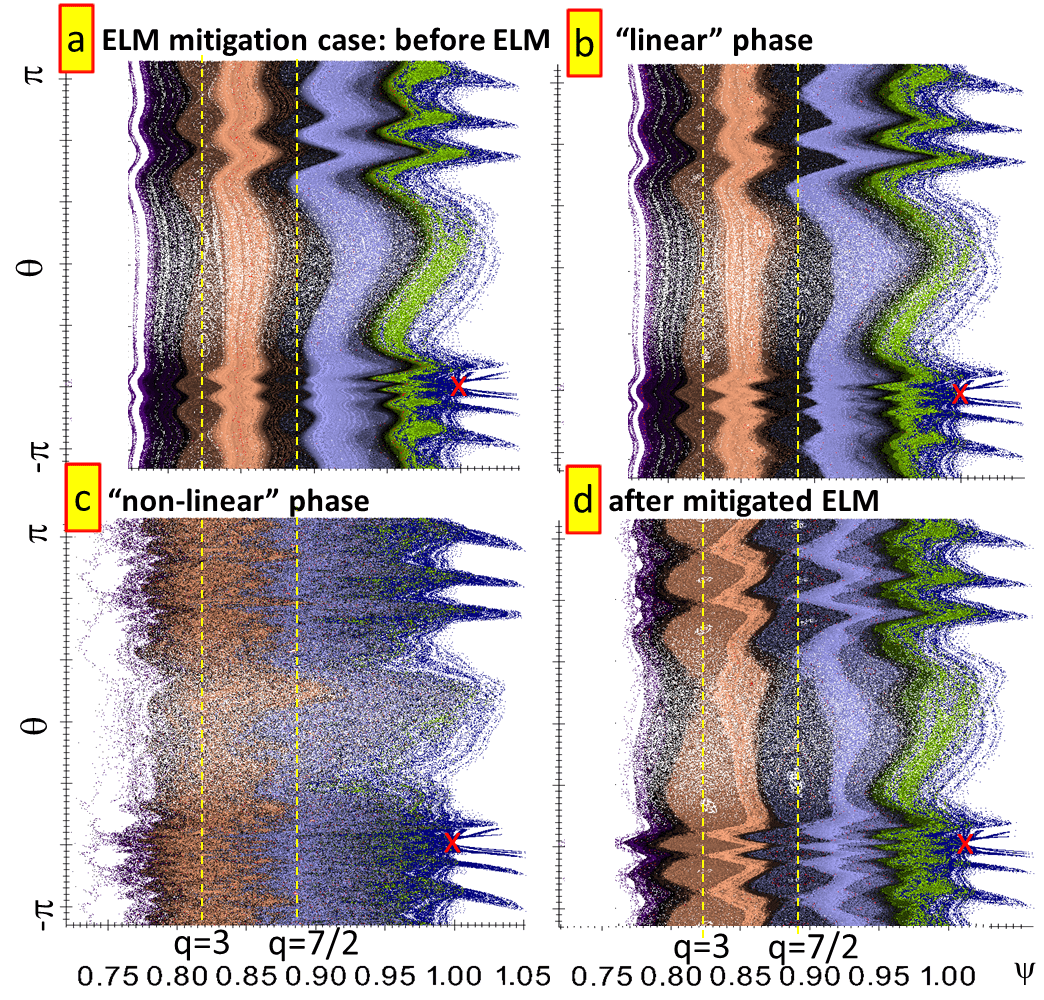}
\caption{Poincar\'e plot of the magnetic topology in the ELM mitigation case, in $[\psi,\theta]$ coordinates: (a) at the time when $n>2$ modes are added in the simulation; (b) in the ``linear" phase of the mitigated ELM; (c) in its non-linear phase; (d) after the mitigated ELM. The arbitrary colors representing the connection length allow to highlight the magnetic surfaces. The $q=3$ and $q=7/2$ surfaces are emphasized by yellow lines. The X-point location is underlined by the red crosses.}
\label{fig:fig11}
\end{figure}

\begin{figure}[h!]
\centering
\includegraphics[width=\textwidth]{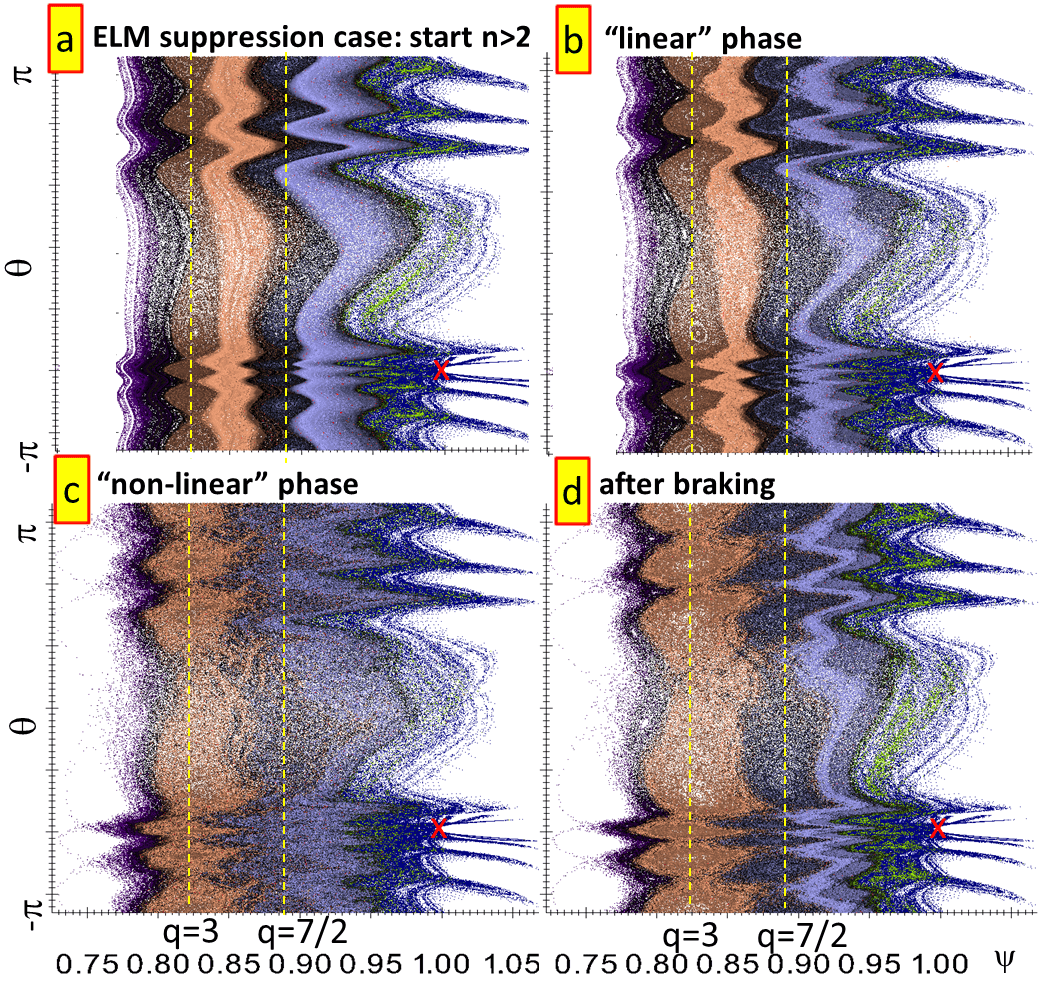}
\caption{Poincar\'e plot of the magnetic topology in the ELM suppression case, in $[\psi,\theta]$ coordinates: (a) at the time when $n>2$ modes are added in the simulation; (b) in the ``linear" phase of the suppressed ELM; (c) in its non-linear phase; (d) after the magnetic activity. The arbitrary colors representing the connection length allow to highlight the magnetic surfaces. The $q=3$ and $q=7/2$ surfaces are emphasized by yellow lines. The X-point location is underlined by the red crosses.}
\label{fig:fig12}
\end{figure}

The evolution of the magnetic topology at the edge in ELM mitigation and ELM suppression cases is plotted respectively in Figs. \ref{fig:fig11} and \ref{fig:fig12}, where Poincar\'e plots are presented for four different times: (a) just before the inclusion of $n \geq 4$ in simulation, so for the 3D-equilibrium induced by $n=2$ RMPs, at $t=0.6ms$ ; (b) during the ``linear" growth phase of the modes $n \geq 4$, at $t \approx 0.8ms$ ; (c) during the ``non-linear" phase when the mode-mixing is close to the maximum, at $t \approx 1.6ms$ ; (d) after the mitigated or suppressed ELM regime, at $t \approx 2ms$. In these Poincar\'e plots, the connection length of the field lines allows the different field line structures to be highlighted. The dark brown structure around $\psi=0.82$ underlines the $q=3$ surface, and the black structure at the pedestal top ($\psi \approx 0.88$) exhibits the $q=7/2$ surface. 
Before including the modes $n \geq 4$ in the simulation (a), a similar $n=2$ structure induced by RMPs is observed in both cases: an ($m=6, n=2$) magnetic island chain appears on the $q=3$ surface and very small islands are formed on the $q=7/2$ surface. A stochastic layer is observed at the very edge, and the kinking of the field lines is maximal near the X-point. Logically, since a larger magnetic perturbation is applied in the ELM suppression case, the magnetic islands and the ergodic layer are larger than in the ELM mitigation case, and the kinking is more important. 
During the ``linear" phase (b), the modes $n=4, 6$ and $8$ are still too small to modify significantly the magnetic structure observed in (a).

In the ``non-linear" phase of the ELM mitigation case (Fig. \ref{fig:fig11}(c)), a complete mixing of the field lines is observed; in particular, the black region around $q=7/2$ is completely blended with the brown and indigo regions, and no more magnetic island structure is observable. This is due to a complete stochastization of the edge from $\psi =0.8$, similarly to the natural ELM case without RMPs. However, in the ``non-linear" phase of the ELM suppression case (Fig. \ref{fig:fig12}(c)), the stochasticity is reduced as compared to the ELM mitigation case: magnetic islands are still discernible on $q=3$ and the black region around $q=7/2$ is not fully melded with the adjacent radial regions. It means that the saturation of the edge-localized modes at a low level, due to the coupling with the $n=2$ RMP-induced mode, reduces the stochasticity level and therefore prevents the edge relaxation from occurring. 

Just after the magnetic activity induced by the ELM mitigation or ELM suppression regimes (d), the stochasticity level has decreased and magnetic islands appear again on $q=6$ and $q=7/2$, with a larger size than in the initial phase (a). These magnetic islands are larger in the ELM suppression case than in the ELM mitigation case, since the RMP penetration has been enhanced by the resonant braking. 

\section{Conclusion and discussion} \label{sec:conclu}

\subsection{Discussion}  \label{sec:discu}
A few points described in this paper can be discussed. 
First of all, the enhanced resistivity in the simulations induces larger magnetic islands and a larger chaotic layer at the edge as compared to fully realistic parameters. However, even at lower resistivity, the bifurcation towards a penetrated state of RMPs can occur at the pedestal top if the perpendicular electron flow is weak, which means that magnetic island chains at the pedestal top can be produced in the same way but with a slightly reduced island size. Furthermore, the identification of the rotation braking and island penetration at the pedestal top as ELM suppression criterion in DIII-D and KSTAR (as discussed in \ref{sec:rotation}) supports the mechanisms described by this modeling. 

The other main limitation of this modeling is the reduced set of toroidal mode numbers included in simulation (up to $n=8$). As commented in \ref{sec:elec}, the inclusion of a wide range of mode numbers up to $n=40$ might have allowed us to observe not only the inverse energy transfer from ``medium" $n=6-8$ to lower $n$ modes, but also a direct transfer towards larger $n \geq 30$ modes. As proposed by Refs. \citep{Kim_IAEA18, Kim_NF19}, this direct transfer can induce an enhanced stochasticity and might also explain the large density pumpout and the broadband turbulence observed in ELM suppression regimes of ASDEX Upgrade. Such turbulent simulations including a large number of toroidal harmonics is currently out of reach computationally in realistic tokamak geometry. Nevertheless, the reduction of the simulation to the essential ingredients allows us to describe accurately the mechanism of the ELM suppression.

Besides, the experimental parameters used as input for this modeling are extracted from an ASDEX Upgrade discharge at low triangularity where only the ELM mitigation is obtained, not the ELM suppression. For a nominal RMP amplitude corresponding to the experimental value, the ELM mitigation is also found in modeling. Then the RMP amplitude in modeling was increased to a level which cannot be reached in the experiments, allowing to observe the ELM suppression. Later on, in the experiments, the ELM suppression was obtained in other discharges at high triangularity. It is likely that a large RMP penetration occurs during these high-triangularity discharges, in the same way as it is observed in our modeling when an enhanced RMP amplitude is applied. The possible link between an increased RMP penetration at high triangularity and an enhanced peeling-kink response, disclaimed by 1-fluid linear MHD simulations \cite{Ryan_2019}, should be verified through 2-fluid non-linear MHD modeling in future work.

Last, it is interesting to discuss the different theories currently proposed to explain ELM suppression by RMPs. On the one hand, several works (e.g. Ref. \cite{Wade_NF15}) suggest that ELM suppression is obtained when RMPs sufficiently enhance the edge transport, such that the pressure gradient and current density are reduced below the ELM stability limit at the pedestal top, similarly to the initial goal of RMPs described in introduction. On the other hand, this paper describes an alternative mechanism of ELM suppression, in which the reduction of the pressure gradient and current density below the stability limit is not a necessary condition. In this mechanism, the toroidal coupling between the ELMs and the RMP-induced mode induce the saturation at a low level of the edge modes and the sudden braking of their rotation. This mechanism is supported by experimental observations of DIII-D and KSTAR and does not contradict the observations of several ASDEX Upgrade discharges. However in other ASDEX Upgrade experiments \cite{Suttrop_NF18}, the electron perpendicular flow does not cross zero at the pedestal top, which suggests that another ELM suppression mechanism or regime may exist. A kinetic resonance where the $E \times B$ velocity is zero is one candidate to explain this other ELM suppression regime \cite{Heyn_NF14}. From our modeling, one can propose another possible explanation: when RMPs amplify edge peeling-kink modes, the toroidal coupling between modes forces the peeling-ballooning modes to saturate at a low level; if the saturation level is low enough, the peeling-ballooning modes could be replaced by the saturated peeling-kink modes amplified by RMPs at the edge, in an analogous way as the Edge Harmonic Oscillation in the QH-mode regime \cite{Snyder_NF07}. 
Finally, the resonant window for which the ELM suppression is observed in a narrow range of safety factor values $\Delta q$ could be explained by two different things. Either the ELM suppression is operational in a narrow $\Delta q$ range, in which the position of a resonant surface at the pedestal top matches with a weak electron perpendicular flow. Or the amplification of the edge peeling-kink modes, which depends on the alignment of the RMPs with the edge magnetic surfaces and thus on the edge $q$ profile, is strong enough to induce the ELM suppression in a narrow $\Delta q$ range. Further work will aim at testing these possibilities in order to propose a reliable ELM-suppression criterion for ITER. The transition from L- to H- mode while applying RMPs should also be addressed in future modeling since it is essential for ITER to develop scenarii in which the plasma is ELM-free during the complete discharge.

\subsection{Conclusion}
The interaction between ELMs and RMPs was modeled with the magnetohydrodynamic code JOREK using ASDEX Upgrade experimental data as input. The ELM mitigation or suppression is obtained when the applied magnetic perturbation is aligned with the edge field lines. In this case, the poloidal mode coupling between the peeling-kink modes amplified by RMPs and the tearing modes induced by RMPs allow for the penetration of the magnetic perturbations. We therefore call this configuration ``resonant condition". In non-resonant configuration, the applied magnetic perturbation is strongly screened and therefore has almost no effect on the ELM relaxation. In resonant configuration, the ELM mitigation or suppression is induced by the toroidal coupling of the edge-localized peeling-ballooning modes with the $n=2$ perturbation induced by RMPs, which forces the edge modes to saturate at a lower level. If the coupling is relatively low (for a moderate applied RMP amplitude), the saturated edge modes are still rotating and induce a small relaxation, corresponding to the ELM mitigation. If the coupling is strong enough (above a threshold in applied RMP amplitude), the edge modes suddenly slow down until they are locked to RMPs and become static. In this case, the ELM relaxation is fully suppressed. The mode braking is induced by the bifurcation towards the complete penetration of the RMPs inducing the resonant braking of the plasma rotation on the resonant surfaces at the pedestal top. Phenomenologically, the ELM relaxation is characterized by the transient suppression of the radial electric field and a full stochastization of the magnetic field at the edge. In the ELM suppression regime, the electric field does not vanish and the stochasticity level is reduced by the saturation of the edge modes at a low level. The description of the mechanism of the ELM suppression above a threshold in RMP amplitude is an important step towards the definition of an ELM-suppression criterion for ITER.

\subsection*{Acknowledgments}

\textsl{This work has been carried out within the framework of the EUROfusion Consortium and has received funding from the Euratom research and training programme 2014-2018 and 2019-2020 under grant agreement No 633053. Part of this work was carried out using the Marconi-fusion supercomputer operated at Cineca, Italy, and the HELIOS supercomputer system at Computational Situational Centre of International Fusion Energy Research Centre (IFERC-CSC), Aomori, Japan, under the Broader Approach collaboration between Euratom and Japan, implemented by Fusion for Energy and JAEA. The first author acknowledges interesting discussions with Timoth\'ee Nicolas, Pascale Hennequin, Marco Cavedon, Hinrich L\"utjens, \'Elis\'ee Trier and Vinodh Bandaru. The views and opinions expressed herein do not necessarily reflect those of the European Commission.}

\bibliographystyle{unsrt}
\bibliography{biblio_3}

\end{document}